\documentclass[prd,preprint,floatfix,nofootinbib,amsmath]{revtex4}
\usepackage{rotate,subfig}
\usepackage{epsfig}
\usepackage{graphicx}
\usepackage{pstricks}
\usepackage{enumerate}

\catcode`\@=11
\def\sla@#1#2#3#4#5{{%
 \setbox\z@\hbox{$\m@th#4#5$}%
 \setbox\tw@\hbox{$\m@th#4#1$}%
 \dimen4\wd\ifdim\wd\z@<\wd\tw@\tw@\else\z@\fi
 \dimen@\ht\tw@
 \advance\dimen@-\dp\tw@ \advance\dimen@-\ht\z@
 \advance\dimen@\dp\z@
 \divide\dimen@\tw@ \advance\dimen@-#3\ht\tw@
 \advance\dimen@-#3\dp\tw@ \dimen@ii#2\wd\z@
 \raise-\dimen@\hbox to\dimen4{%
 \hss\kern\dimen@ii\box\tw@\kern-\dimen@ii\hss}%
 \llap{\hbox to\dimen4{\hss\box\z@\hss}}}}

\def\slashed#1{%
 \expandafter\ifx\csname sla@\string#1\endcsname\relax
{\mathpalette{\sla@/00}{#1}}
\fi}
\def\declareslashed#1#2#3#4#5{%
 \expandafter\def\csname sla@\string#5\endcsname{%
#1{\mathpalette{\sla@{#2}{#3}{#4}}{#5}}}}
 \catcode`\@=12
\declareslashed{}{/}{.08}{0}{D}
 \declareslashed{}{/}{.1}{0}{A}
 \declareslashed{}{/}{0}{-.05}{k}
 \declareslashed{}{/}{.1}{0}{\partial}
 \declareslashed{}{\not}{-.6}{0}{f}
\def\lsim{\mathrel {\vcenter {\baselineskip 0pt \kern 0pt
    \hbox{$<$} \kern 0pt \hbox{$\sim$} }}}
\def\gsim{\mathrel {\vcenter {\baselineskip 0pt \kern 0pt
    \hbox{$>$} \kern 0pt \hbox{$\sim$} }}}
\def\slashchar#1{\setbox0=\hbox{$#1$}           % set a box for #1
 \dimen0=\wd0                                 % and get its size
  \setbox1=\hbox{/} \dimen1=\wd1               % get size of /
\ifdim\dimen0>\dimen1                        % #1 is bigger
  \rlap{\hbox to \dimen0{\hfil/\hfil}}      % so center / in box
  #1                                        % and print #1
  \else                                        % / is bigger
 \rlap{\hbox to \dimen1{\hfil$#1$\hfil}}   % so center #1
   /                                         % and print /
  \fi}                                         %
\def\cpto{\mathrel {\vcenter {\baselineskip 0pt \kern 0pt
    \hbox{$CP$} \kern 0pt \hbox{$\longrightarrow$} }}}
\def\cptof{\mathrel {\vcenter {\baselineskip 0pt \kern 0pt
    \hbox{$~CP$} \kern 0pt \hbox{$\longleftrightarrow$} }}}

\begin{document}

\baselineskip=15pt
\preprint{}

\title{Flavor changing $Z^\prime$ couplings at the LHC}

\author{Sudhir Kumar Gupta and G. Valencia}

\email{skgupta@iastate.edu,valencia@iastate.edu}

\affiliation{Department of Physics, Iowa State University, Ames, IA 50011.}

\date{\today}

\vskip 1cm
\begin{abstract}

Models with a non-universal $Z^\prime$ exhibit in general flavor changing neutral currents (FCNC) at tree-level. When the $Z^\prime$ couplings favor the third generation, flavor changing transitions of the form $Z^\prime tc$ and  $Z^\prime tu$ could be large enough to be observable at the LHC. In this paper we explore this possibility using  the associated production of a single top-quark with the $Z^\prime$ and find that integrated luminosities of a few hundred fb$^{-1}$ are necessary to probe the interesting region of parameter space.

\end{abstract}

\pacs{PACS numbers: }

\maketitle

\section{Introduction}

One of the most frequently studied extensions of the Standard Model (SM) is an additional $U(1)^\prime$ symmetry and its associated $Z^\prime$ boson \cite{review}. Although the $U(1)^\prime$ charges are family universal in most of the models discussed in the literature, this need not be the case.  
It is well known that a non-universal $Z^\prime$ induces tree-level FCNC which are severely constrained by experiment, most notably meson mixing. The contribution of these FCNC to many processes has been studied in detail in the literature, and generally it has been found that the most severe constraints arise from $K$, $D$, $B_d$ and $B_s$ meson mixing \cite{fcncconstraints}.

It was pointed out recently in Ref.~\cite{He:2009ie}  that there is a class of models, with a down-quark mass matrix of the Georgi-Jarlskog form \cite{Georgi:1979df}, in which the FCNC in the down-type quark sector vanish or are strongly suppressed. These models then satisfy all the existing experimental constraints on the induced FCNC. Interestingly, they also predict that the largest FCNC occurs in the $tc$ transition with a strength comparable to $V_{cb}$. Similarly in these models, the strength of the $Z^\prime tu$ coupling can be  comparable to $V_{ub}$. 

The LHC will serve as a top factory and will thus provide an ideal environment to study this possibility. There exist many studies on $t\to c$ flavor changing transitions at the LHC  \cite{tcgz}, but for the most part they assume that the transition will show up as a non-standard top-quark decay.  This is what happens in scenarios with anomalous $tcg/\gamma/Z$ couplings. However, rare top-quark decay does not produce significant constraints for $tcZ^\prime$ couplings, since the $Z^\prime$ is required phenomenologically  to be heavier than the top-quark \cite{review}. A suitable alternative to cover this case is to  observe the FCNC via single top-quark production.  A possibility that has been discussed in Ref.~\cite{fcnczp}, is the production  
of $t\bar{c}$  and $c\bar{t}$ final states from on-shell $Z^\prime$ production using the largest $tcZ^\prime$ coupling allowed by $D-\bar{D}$ mixing. This procedure suffers from two drawbacks. First, the  cross-section  depends not only on the FCNC transition but also on the flavor diagonal couplings of the $Z^\prime$ to light quarks. This feature makes the predictions very model dependent, a problem that has been addressed in Ref.~\cite{fcnczp} by surveying a large range of models. Second, it appears that the signal will be very difficult to separate from background at the LHC. 

In this paper we discuss a different process, $pp \to t Z^\prime X$, in which the $Z^\prime$ is produced in association with a single top-quark via a FCNC transition. This process has a smaller cross-section  than the one discussed in Ref.~\cite{fcnczp}, but it also has significantly lower background. It can also be described in a more model independent way as we will show. The relevant cross-sections for this process are sufficiently small that experimental studies of this type will not happen early in the LHC program, but rather after a $Z^\prime$ has been discovered and integrated luminosities of several hundred fb$^{-1}$ have been collected. For this reason we only present numerics for $\sqrt{S} = 14$~TeV.

\section{$Z^\prime$ couplings and FCNC}

Our aim is to produce a phenomenological study of $Z^\prime$ FCNC couplings to the top quark at the LHC in a model independent manner. The necessary input to this end consists of the FCNC $Z^\prime$ couplings which we simply parametrize as
\begin{eqnarray}
{\cal L}_{FCNC} = {g\over 2\cos\theta_W} \,
\left( a_{tc} \bar{t}\gamma^\mu P_L c + b_{tc} \bar{t}\gamma^\mu P_R c+
a_{tu} \bar{t}\gamma^\mu P_L u + b_{tu} \bar{t}\gamma^\mu P_R u\right)
Z^{\prime_\mu}\,+{\rm h.c.}.
\label{zpcoup}
\end{eqnarray}

The $Z^\prime$ boson is assumed to be heavier than the top-quark, in keeping with available experimental constraints \cite{review}. In this scenario, rare top-quark decays mediated by a $Z^\prime$ FCNC are very much suppressed and do not play an important role in constraining the couplings in Eq.~\ref{zpcoup} \cite{He:2009ie}. 

FCNC couplings as in Eq.~\ref{zpcoup} arise in models with non-universal $Z^\prime$ couplings and are generally unknown. Typically they are accompanied by corresponding FCNC couplings in the down-quark sector that are severely constrained by meson mixing and rare meson decay observations. Ref.~\cite{He:2009ie} pointed out a mechanism to suppress these FCNC in the down-quark sector to phenomenologically acceptable levels which results in specific predictions for the couplings in Eq.~\ref{zpcoup}. For example, for right handed couplings,
\begin{eqnarray}
b_{tc} \sim  V_{cb} \sim {\cal O}(5\times 10^{-2}) &&
b_{tu} \sim  V_{ub} \sim {\cal O}(5 \times 10^{-3}).
\label{expsizeR}
\end{eqnarray}
In Ref.~\cite{fcnczp} on the other hand, it is assumed that the FCNC couplings are left handed and they are directly constrained by $D-\bar{D}$ mixing resulting in similar numbers,
\begin{eqnarray}
a_{tc} \sim  V_{cb} \sim {\cal O}(5\times 10^{-2}) &&
a_{tu} \sim  V_{ub} \sim {\cal O}(5 \times 10^{-3}).
\label{expsizeL}
\end{eqnarray}
These numbers represent a benchmark of sensitivity to FCNC needed to be competitive with existing indirect constraints.

The production cross-section for the process we discuss in this paper, $pp \to t {\rm ~(or~}\bar{t}) Z^\prime X$, depends only on SM parameters, the couplings in Eq.~\ref{zpcoup}, and the $Z^\prime$ mass. Additional model dependence arises in the different flavor diagonal couplings of the $Z^\prime$. In our process, these additional couplings enter only to determine the final states (and their corresponding branching ratio) in which the $Z^\prime$ is observed. To keep our study as model independent as possible we simply present our results in terms of these branching ratios.

\section{LHC phenomenology}

\subsection{Preliminaries}

We now discuss single top quark production in association with the $Z^\prime$, $pp\to t ({\rm or ~}\bar{t})Z^\prime X$, at the LHC. In models with non-universal $Z^\prime$ bosons that result in FCNC, the flavor diagonal couplings of the $Z^\prime$ to the third generation are usually preferred. With this in mind we consider several possibilities for the observation of the $Z^\prime$ in this channel. Our first scenario, that will serve as a benchmark for the feasibility of these studies, will assume that the $Z^\prime$ decays into muon pairs with non-negligible branching ratio 
${\cal B}_{Z^\prime \to \mu^+\mu^-}$. To allow for the possibility of  enhanced branching ratios to third generation fermions we next consider  $Z^\prime$ decays into tau-lepton pairs with non-negligible branching ratio 
${\cal B}_{Z^\prime \to \tau^+\tau^-}$. Finally, to include the possibility of a leptophobic $Z^\prime$ we also discuss the cases in which the $Z^\prime$ decays into $b\bar{b}$ or $t\bar t$ pairs with non-negligible branching ratio.

\begin{figure}
\centering
\includegraphics[angle=-90,width=.49\textwidth]{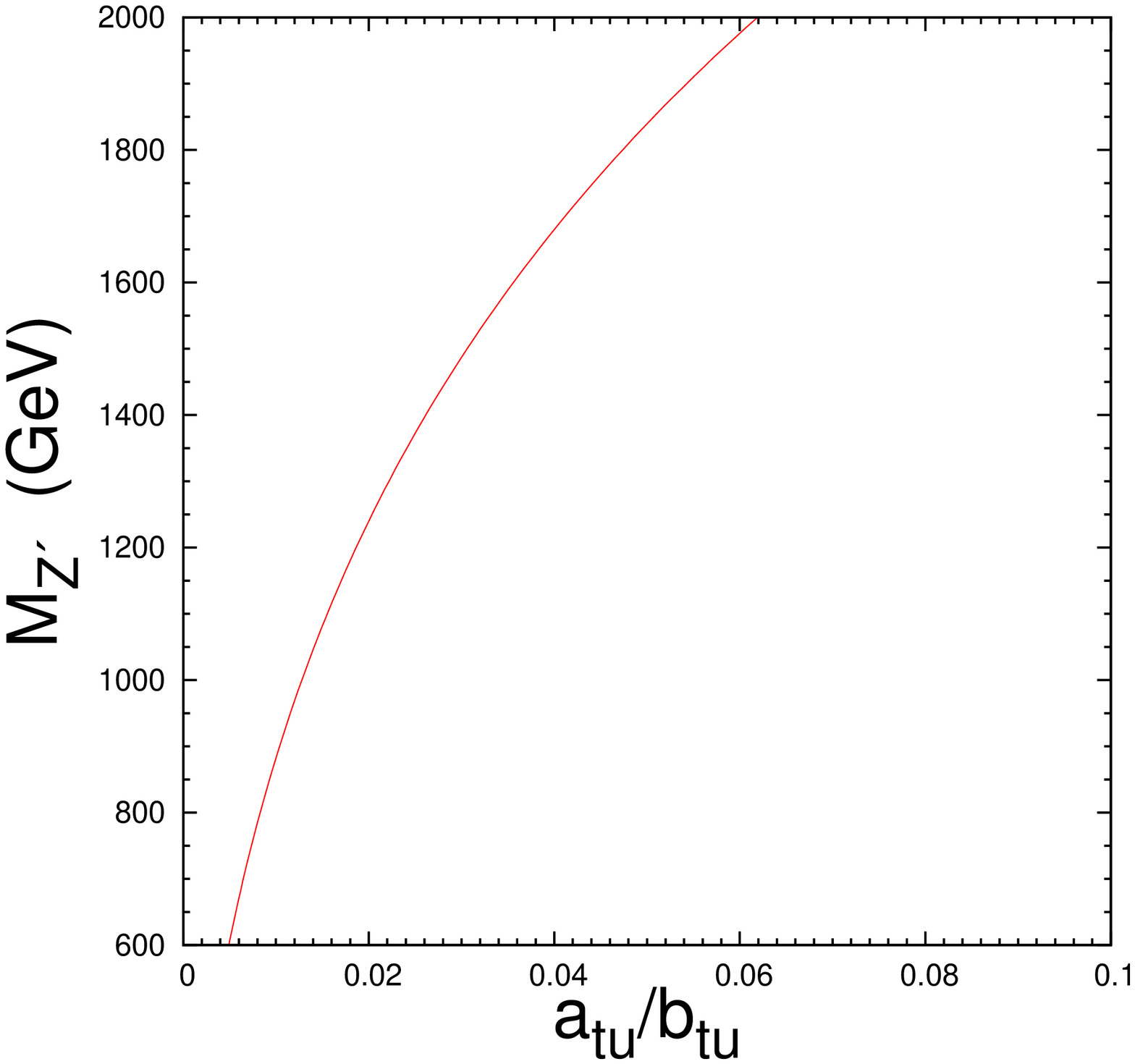}
\includegraphics[angle=-90,width=.49\textwidth]{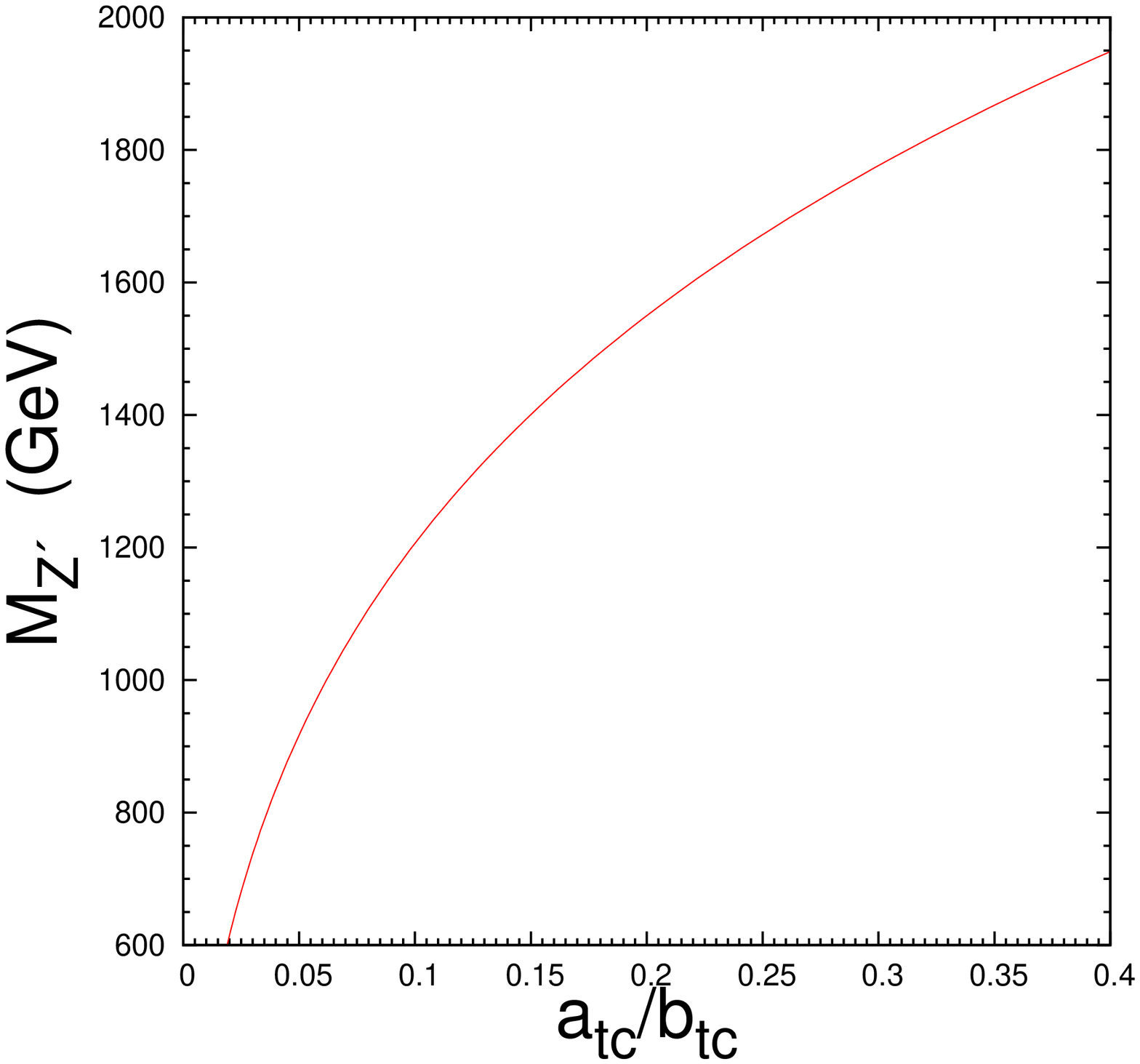}
\caption{The region to the right of the curve shows the parameter space yielding more than 10 $(tZ^\prime +\bar{t}Z^\prime)$ events at the LHC  for $\sqrt{s} = 14$ TeV and $\int {\cal L} dt = 100 fb^{-1}$.} 
\label{f:prelim}
\end{figure}

In order to gain some insight into the numerology involved, we first consider $pp \to  Z^\prime t {\rm ~(or~}\bar{t})$ disregarding background and identification efficiencies. We use Madgraph \cite{madgraph} and check our results with Comphep \cite{comphep} for this preliminary calculation by implementing the new couplings of Eq.~\ref{zpcoup} into both programs. The signal arises from the parton subprocesses $cg\to Z^\prime t$ or $ug\to Z^\prime t$ (or those with $\bar{t}$ instead of $t$) for the $a_{tc}, b_{tc}$ or $a_{tu},b_{tu}$ couplings respectively. Furthermore, since there is no interference between the new couplings and any standard model process, the resulting cross-sections are quadratic in the new couplings and identical for the left and right handed cases. The $utZ^\prime$ couplings are expected to be smaller than the $ctZ^\prime$ in both the cases discussed in Ref.~\cite{He:2009ie} and Ref.~\cite{fcnczp} by about an order of magnitude as seen in Eqs.~\ref{expsizeR},~\ref{expsizeL}. This, however, is compensated by the larger up-quark parton distribution functions that enter the $ug\to Z^\prime t$ process. Using the CTEQ-6L1 parton distribution functions, $\sqrt{S}=14$~TeV, and $m_t = 174.3$~GeV we obtain the following raw cross-sections  for two values of the $Z^\prime$ mass:
\begin{eqnarray}
\sigma(pp\to tZ^\prime +\bar{t}Z^\prime) &\approx& 
\left( 0.6 \ a_{tc}^2 + 7.8 \ a_{tu}^2 \right){\rm ~pb},\  {\rm ~for~}M_{Z^\prime}=500~{\rm GeV} \nonumber \\
\sigma(pp\to tZ^\prime +\bar{t}Z^\prime) &\approx& 
\left( 0.03 \ a_{tc}^2 + 0.7 \ a_{tu}^2 \right){\rm ~pb},\  {\rm ~for~}M_{Z^\prime}=1~{\rm TeV},
\label{roughestimate}
\end{eqnarray}
and identical numbers for right handed couplings with $b_{tc},b_{tu}$ replacing $a_{tc},a_{tu}$. These cross-sections can be turned into a rough lower bound for the FCNC that would be observable at LHC by arbitrarily demanding a signal yield of 10 events for an integrated luminosity of 100 fb$^{\-1}$. The resulting numbers are shown in Table~\ref{tabrough}.
\begin{table}[h]
\centering
\begin{tabular}{|c|c|}\hline
$M_{Z^\prime}$ (GeV) & Lower limit \\\hline
500 & $g_{tc} > 1\times 10^{-2}$ \\
500 & $g_{tu} > 4\times 10^{-3}$ \\
1000 & $g_{tc} > 6\times 10^{-2}$ \\
1000 & $g_{tu} > 1\times 10^{-2}$  
\\\hline
\end{tabular}
\caption{Lowest value of the respective FCNC coupling ($g=a,b$) that produces 10 $(tZ^\prime +\bar{t}Z^\prime)$ events at LHC with 100 fb$^{-1}$ for a given $M_{Z^\prime}$. }
\label{tabrough}
\end{table}
In Figure~\ref{f:prelim} we show these results as functions of $M_{Z^\prime}$.
%--------------------------------------------------------------------
%--------------------------------------------------------------------

Comparing the numbers in Table~\ref{tabrough} and Figure~\ref{f:prelim} with those in Eqs.~\ref{expsizeR},~\ref{expsizeL}, shows that at least 100 fb$^{\-1}$ will be necessary to place significant constraints on the couplings of Eq.~\ref{zpcoup}. Already at this level one can see that the sensitivity to the $g_{tc}$ couplings is severely reduced relative to the $g_{tu}$ couplings by the relatively small charm content of the proton. This suggests that a higher order process contributing to $pp \to tZ^\prime X$ might be equally or more important than $cg\to Z^\prime t$. Indeed we find that gluon fusion processes such as $gg \to t Z^\prime \bar{c}$ contribute comparable cross-sections.

Armed with these preliminary numbers we proceed to a more detailed numerical study that addresses both background and reconstruction issues upto some extent.  To this end we use Madgraph 
\cite{madgraph} for generation of both signal and background events. To identify the process of interest we consider only semileptonic decay of the top-quark $t\to b \mu \nu$ and four different decay channels for the $Z^\prime$: 
$\mu^+\mu^-$, $\tau^+\tau^-$, $b\bar{b}$ and $t\bar{t}$. For the $\tau^+\tau^-$ and $t\bar{t}$ channels we consider the muonic decay of all tau leptons and top-quarks. Our basic acceptance cuts for the events will be  
\begin{eqnarray}
p_{T_b},\, p_{T_{\mu}}&>& 25 ~{\rm GeV} \nonumber \\
|\eta_\mu|,\, |\eta_b| & \leq & 2.7 \nonumber \\
\delta {R}_{ij} &>& 0.4; \, i,j=\mu,b.
\end{eqnarray}

\subsection{Topologies and Background}

We consider the following decays modes of $Z'$:
(a) $Z'\to \mu^+\mu^-$, (b) $Z'\to \tau^+\tau^- \to \mu^+\mu^-+X$, (c) $Z'\to t \bar{t}\to b\bar{b}\mu^+\mu^-+X$, and, (d) $Z'\to b\bar{b}$ and their corresponding backgrounds at the LHC. 
For the production process $tZ'+\bar{t}Z'$, we thus have the following 
final state topologies (for leptonic decay of top and anti-top) and standard model background:

\begin{enumerate}[a)]

\item $b \mu^+\nu_{\mu}\mu^+\mu^-$ and 
${\bar b} \mu^-{\bar{\nu}}_{\mu} \mu^+\mu^-$ or $b + 3\mu + X$.

\item
$b \mu^+\nu_{\mu}\mu^+\nu_{\mu}{\bar{\nu}}_{\tau}
\mu^-{\bar{\nu}}_{\mu}\nu_{\tau}$ and ${\bar b} \mu^-\nu_{\mu}\mu^+\nu_{\mu}{\bar{\nu}}_{\tau} \mu^-{\bar{\nu}}_{\mu}\nu_{\tau}$
 or $b + 3\mu + X$. 
 
The SM background for cases (a) and (b) is due to the following final states: $\mu^+\mu^+\mu^-{\bar{\nu}}_{\mu}j$, and, $\mu^-\mu^+\mu^-\nu_{\mu}j$ (where $j$ stands for a gluon or any quark and results in one jet). There are 528 SM Feynman diagrams that result in this final state, predominantly they correspond to $W^{\pm} \gamma^{\star}/Z j$ production followed by $W^{\pm}$, $\gamma^{\star}/Z$ decay into muons. The net cross-section for these processes, calculated by Madgraph, is $5.1\times10^{-2}$ pb. This results in  5100 background events for an integrated luminosity of 100 $fb^{-1}$.

We did not consider processes yielding $3\mu + j + 3\nu$, which can also be part of the background to case (b) because this is  expected to be much smaller than the background described above and is much harder to estimate.

\item $b \mu^+\nu_{\mu}b \mu^+\nu_{\mu}{\bar{b}}\mu^-{\bar{\nu}}_{\mu}$ and
${\bar b} \mu^-{\bar{\nu}}_{\mu}{\bar{b}}\mu^-{\bar{\nu}}_{\mu}b \mu^+\nu_{\mu}$ or 
$3b + 3\mu + X$.

A possible background is $t\bar{t}t\bar{b}$ and $t\bar{t}\bar{t}b$, where all the top and anti-top
quarks decay into muons resulting in a final state with $4b+3\mu + 3\nu's$. Of these, those final states in which only three of the four $b$ jets pass the basic acceptance cuts constitute the background. 
We estimate the cross-section for $4b+3\mu+3\nu$ to be about $1.8\times10^{-6}$ pb resulting in negligible background.

\item $b \mu^+\nu_{\mu}b\bar{b}$ and
${\bar b} \mu^-{\bar{\nu}}_{\mu}b\bar{b}$ or $3b + \mu + X$.

In this case the signal consists of three $b$ jets and one muon so the background is mostly $WW j \to \mu \nu + 3j$ and $W Z j\to \mu \nu + 3j$. The cross-sections for these processes are 3.48 pb and 1.83 pb respectively, yielding a total $5.31\times10^5$ 
background events for an integrated luminosity of $100 fb^{-1}$.

\end{enumerate}

\subsection{Cuts}

In order to reduce the background the following sets of selection cuts for cases (a) and (b) were implemented.

\begin{enumerate}[C1]

\item Missing $E_T > 30$ GeV. This will ensure selection of only those final states which have missing energy due to at least one neutrino.

\item Muon pair invariant mass. $m_{\mu_i \mu_j} > 20$ GeV and $|m_{\mu_i\mu_j} - m_Z| > 25$ GeV with $i, j = 
{1,2,3}; i\not= j$  
to eliminate dimuons from photons and $Z$ bosons. Muons are 
$p_T$ ordered i.e. $p_{T_{\mu_1}} > p_{T_{\mu_2}} > p_{T_{\mu_3}}$. 

\item Muon transverse momentum. $p_{T_{\mu_1}}>120$ GeV, $p_{T_{\mu_2}}> 100$ GeV, and, $p_{T_{\mu_3}} > 30$ GeV. This, in addition to reducing the background, will further ensure that the first two muons are indeed due to the decay of a heavy resonance.

\item Scalar sum of transverse momenta of all visible final state 
particles (all jets and leptons). $\Sigma P_{T_{visible}} > 350$ GeV to ensure selection of only high centre-of-mass energy events. 

Applying all these cuts reduces the background (cases (a) and (b)) to 17 events for an 
integrated lumionosity of $100 fb^{-1}$. Table~\ref{effc} shows the efficiency of each these cuts for reducing background and its corresponding effect on the signal for two different values of $M_{Z'}$.

For case (c) we apply the sets of cuts C1 and C2, to ensure that the first two muons and b-jets are indeed due to cascade decays of $Z'$ via top or anti-top quark. In this case, the selection cuts completely eliminate the background. 

Finally, for case (d) we impose the  cuts C1 and C4 as well as a requirement high $b\bar{b}$ pair invariant mass (C5) to remove QCD background.

\item  Large $b\bar{b}$ invariant mass: $m_{b_1b_2} > 300$ GeV.

\end{enumerate}

%%%%%%%%%%%%
\begin{table}[h]
\centering
\begin{tabular}{|c|c|c|c|c|}\hline
cuts& $m_{Z'}=0.6$~ TeV & $m_{Z'}=1$ ~TeV & $m_{Z'}=1.5$ TeV &SM background\\\hline
Basic&57.8&62.3&65.5&45.2\\
Basic + C1&45.5&50.2&52.9&40.8\\
Basic + C1 + C2  &38.6&46.1&50.1&1.1\\
Basic + C1 + C2 + C3 &34.9&42.8&46.7&0.4\\
Basic + C1 + C2 + C3 + C4 &34.8&42.8&46.6&0.3\\\hline
\end{tabular}
\caption{Different cut efficiency (in per cent) in reducing background and its effect on the signal for selected values of $M_{Z^\prime}$ for the case (a) discussed in the text.}
\label{effc}
\end{table}
 
\subsection{Results}

\begin{figure}
\centering
\includegraphics[angle=-90,width=.49\textwidth]{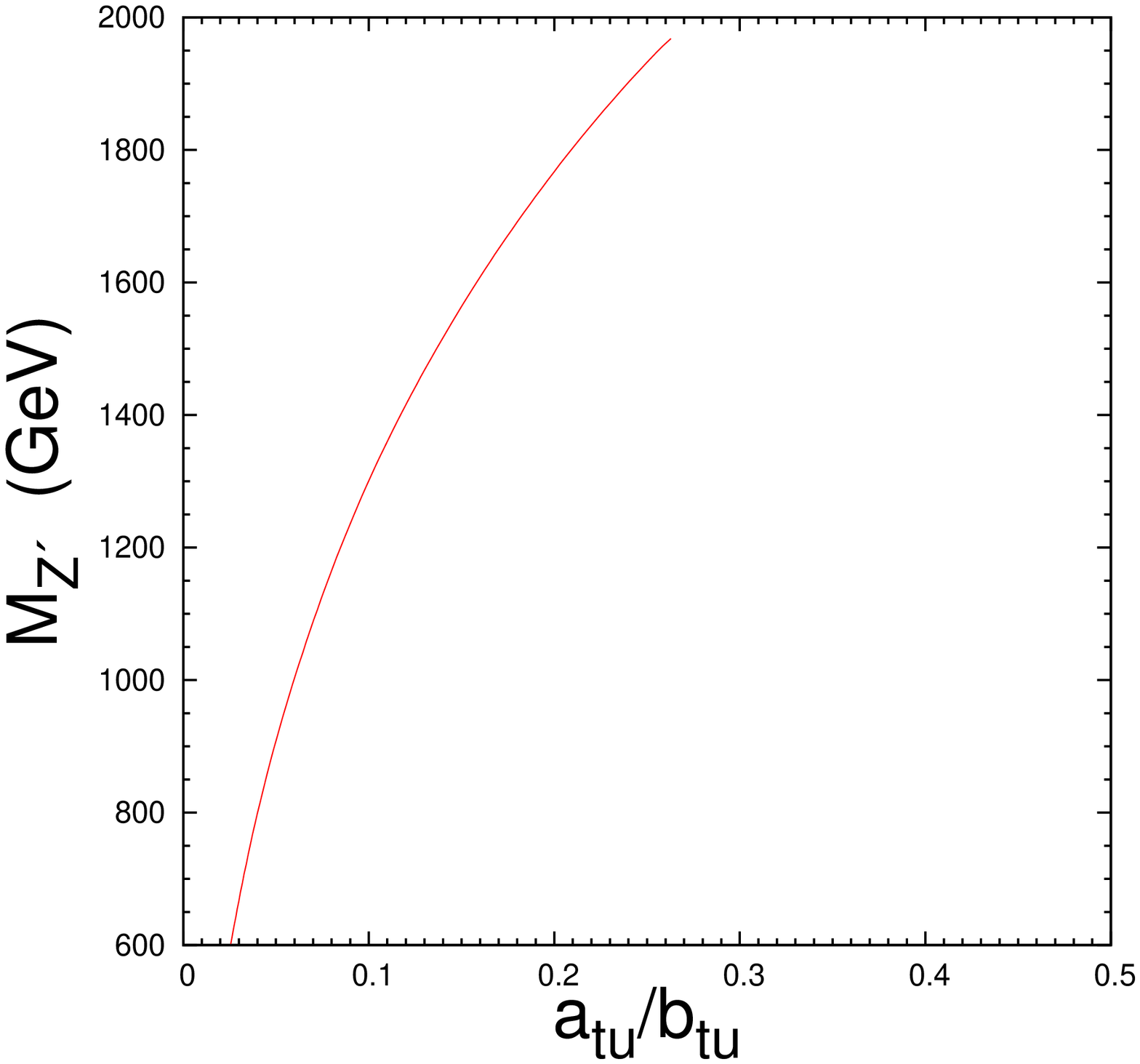}
\includegraphics[angle=-90,width=.49\textwidth]{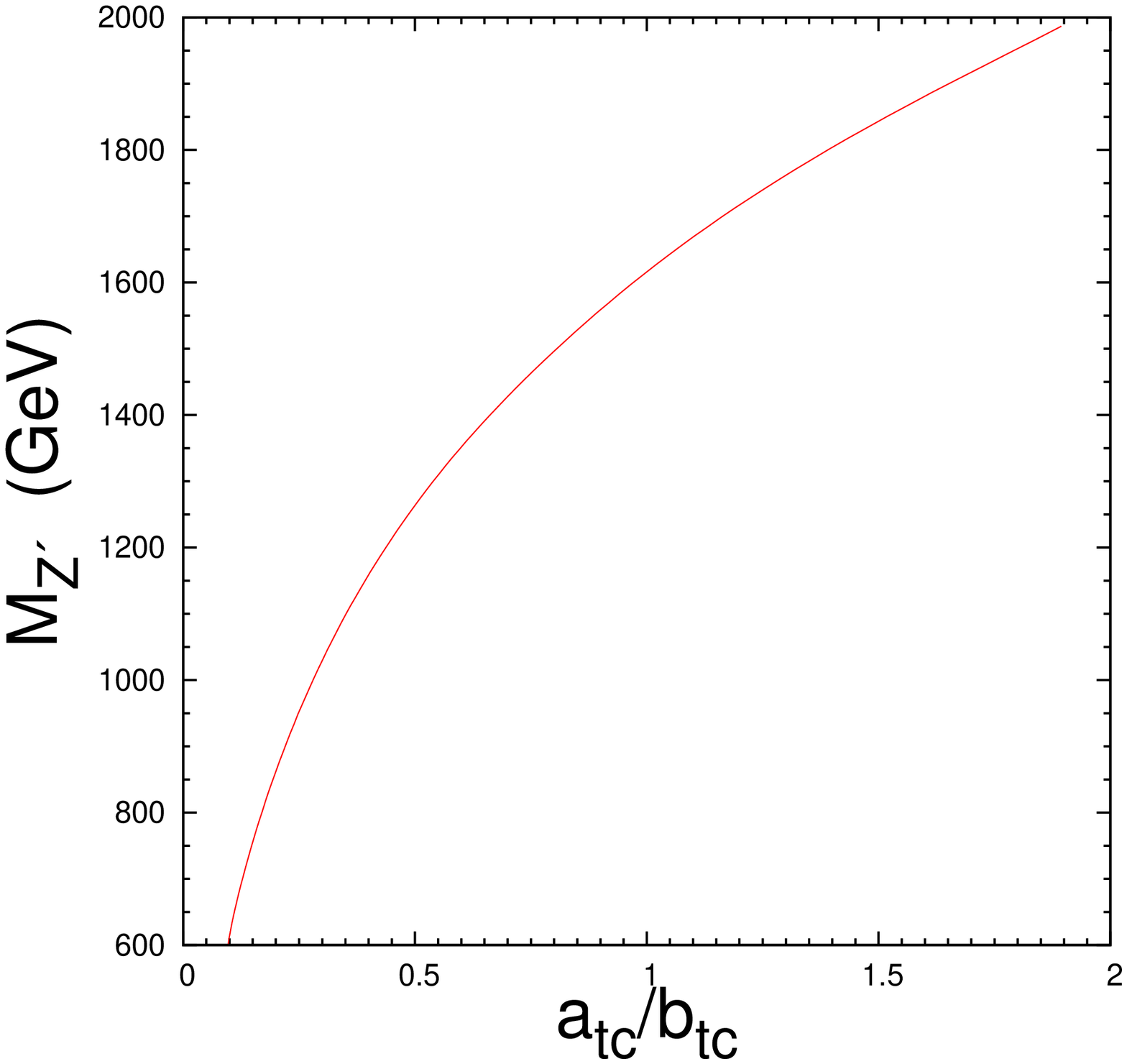}
\caption{The region to the right of the curve shows the parameter space accessible at the LHC in $Z'\to \mu^+\mu^-$ detection mode for $\sqrt{s} = 14$ TeV and $\int {\cal L} dt = 100 fb^{-1}$ after the cuts discussed in the text.} 
\label{f:gmm}
\end{figure}

\begin{figure}
\centering
\includegraphics[angle=-90,width=.49\textwidth]{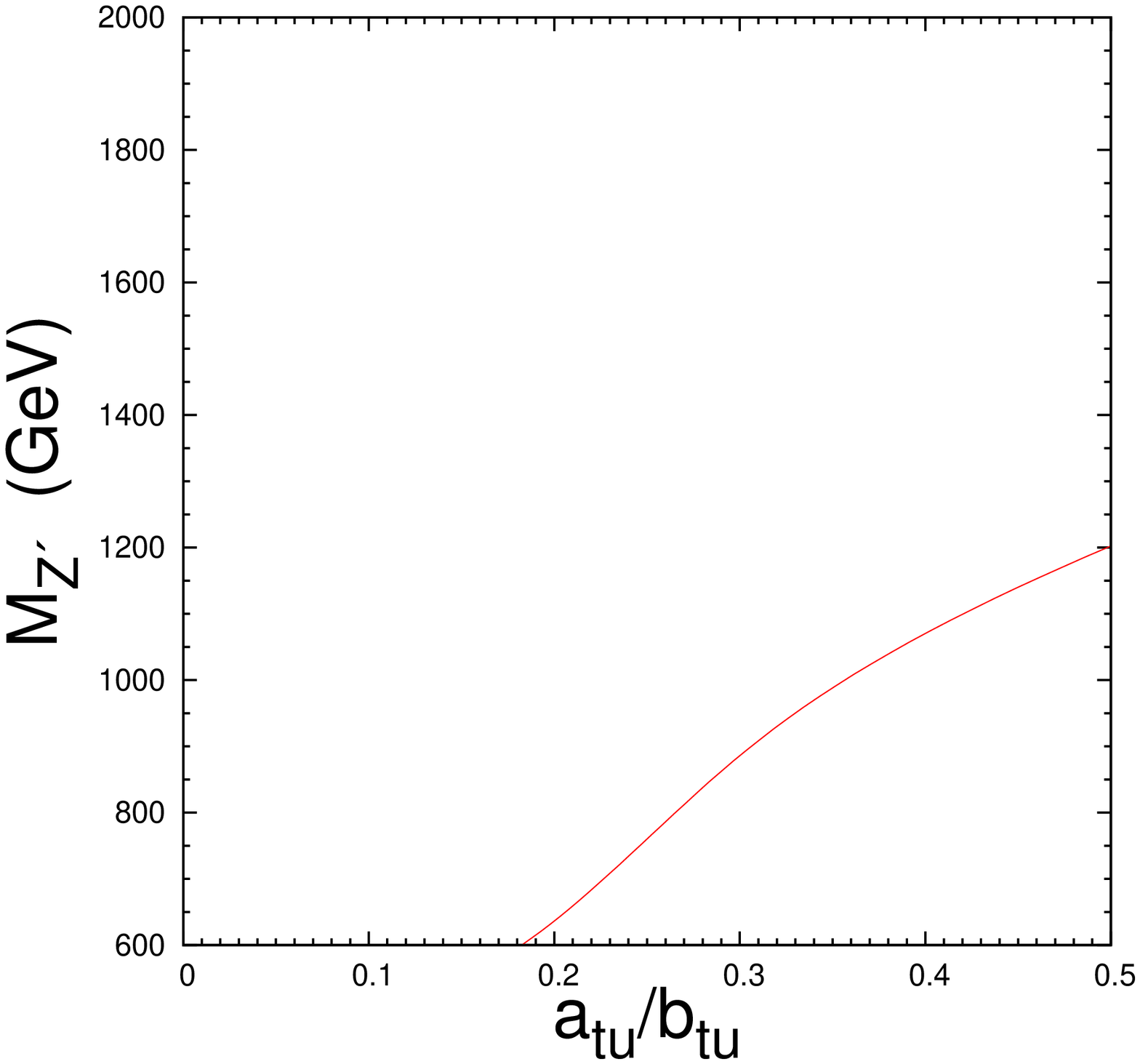}
\includegraphics[angle=-90,width=.49\textwidth]{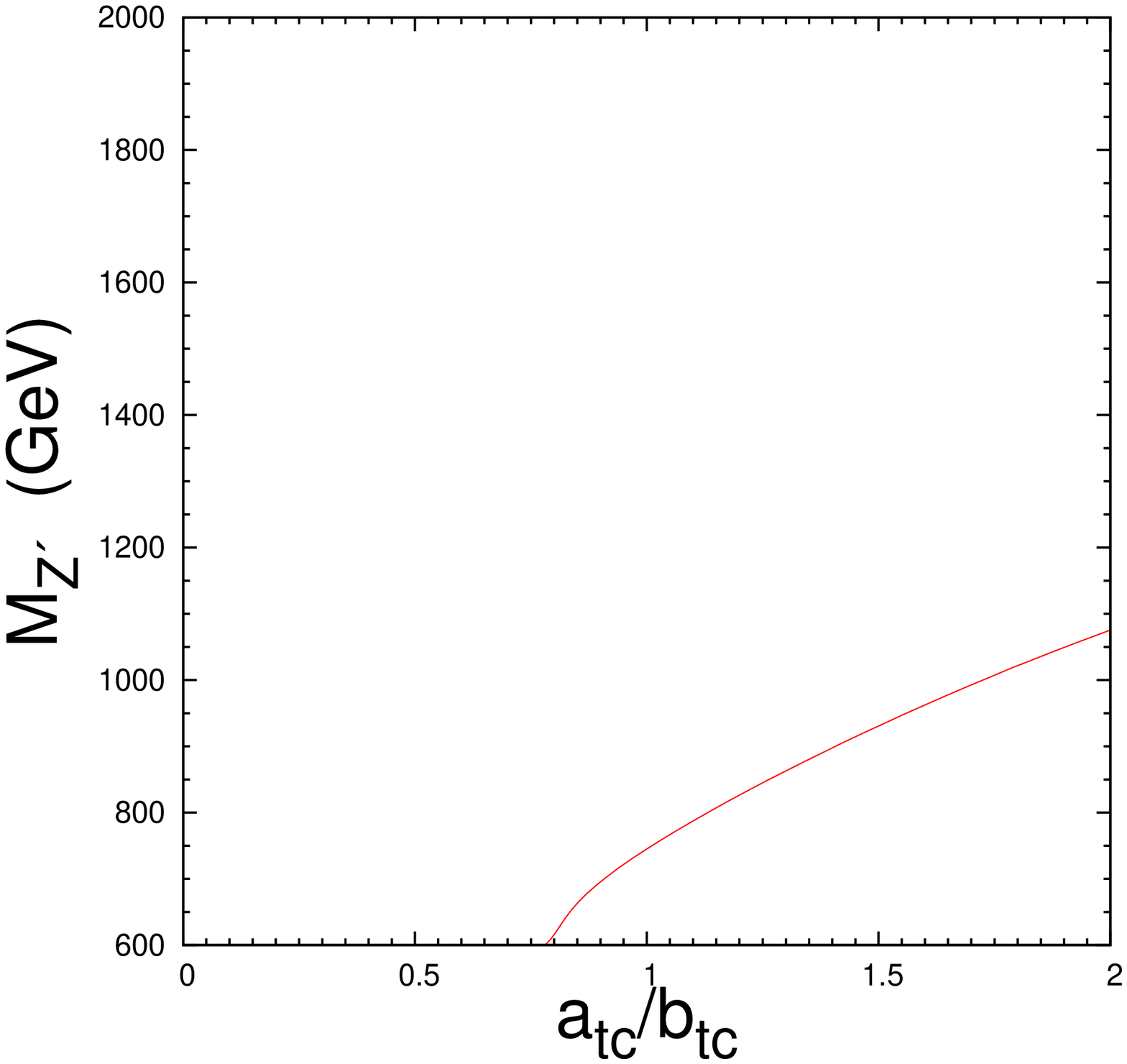}
\caption{The region to the right of the curve shows the parameter space accessible at the LHC in $Z'\to \tau^+\tau^-$ detection mode for $\sqrt{s} = 14$ TeV and $\int {\cal L} dt = 100 fb^{-1}$ after the cuts discussed in the text.}
\label{f:gll}
\end{figure}

\begin{figure}
\centering
\includegraphics[angle=-90,width=.49\textwidth]{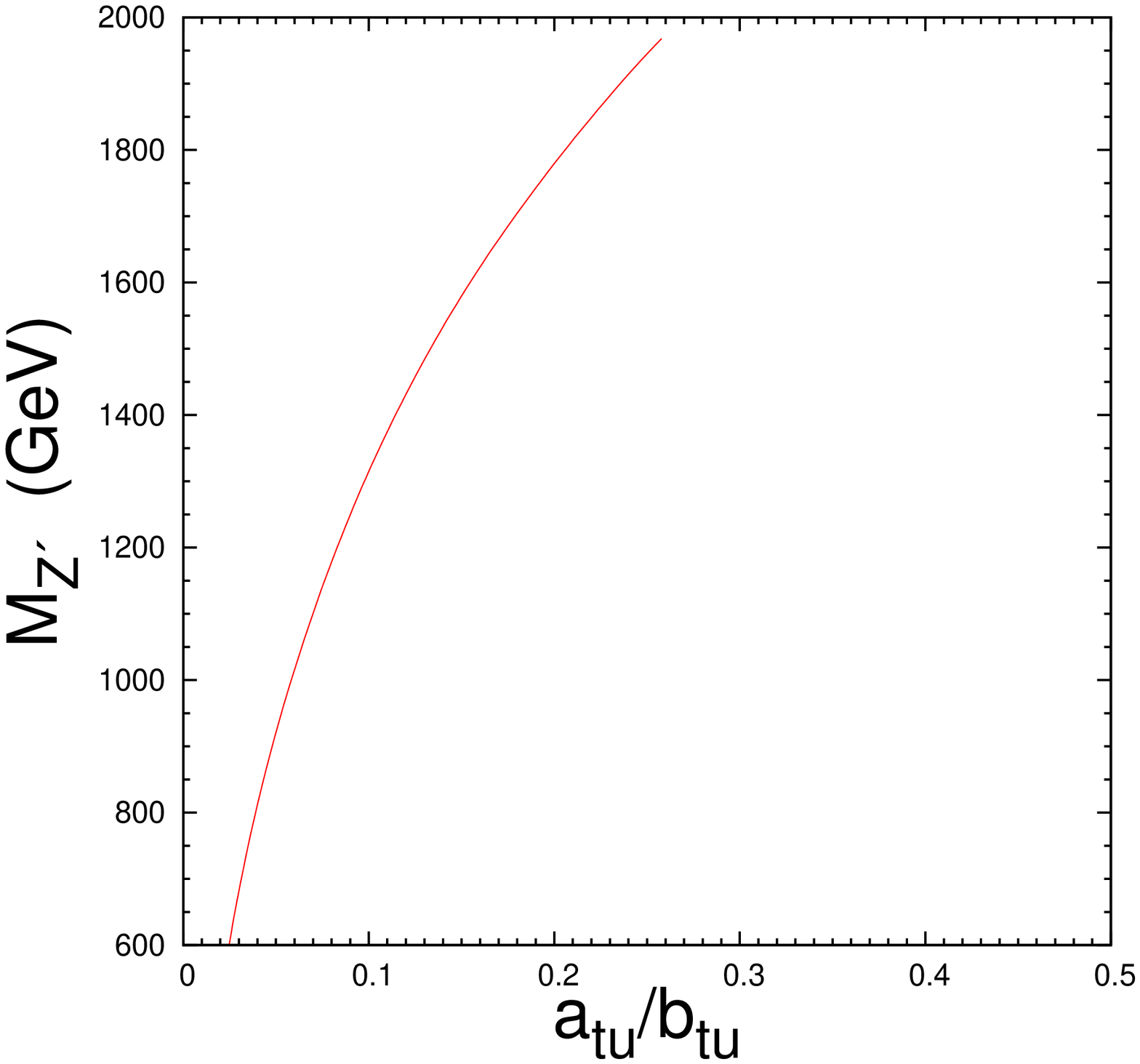}
\includegraphics[angle=-90,width=.49\textwidth]{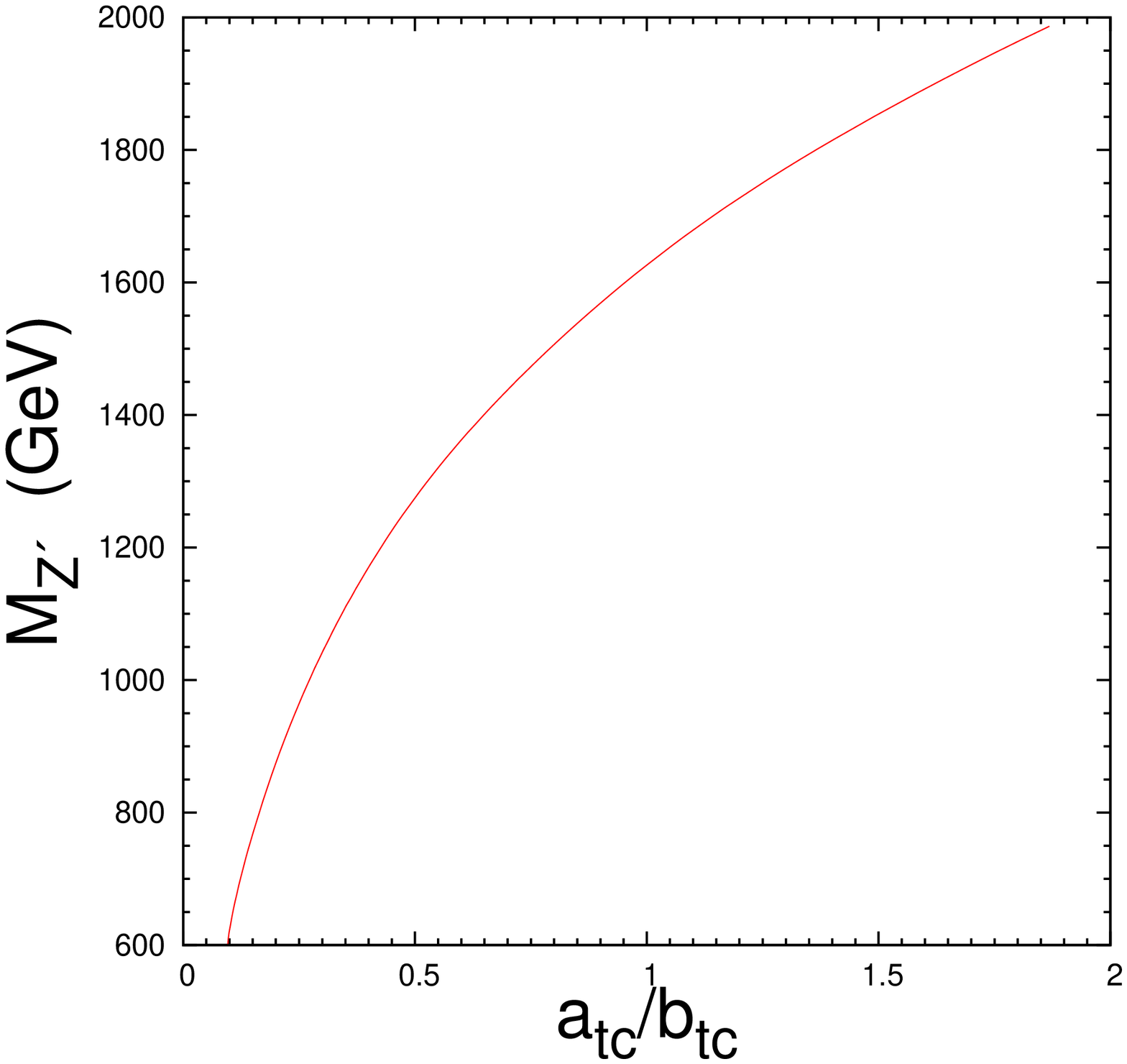}
\caption{The region to the right of the curve shows the parameter space accessible at the LHC in $Z'\to b\bar{b}$ detection mode for $\sqrt{s} = 14$ TeV and $\int {\cal L} dt = 100 fb^{-1}$ after the cuts discussed in the text.}
\label{f:gbb}
\end{figure}

\begin{figure}
\centering
\includegraphics[angle=-90,width=.49\textwidth]{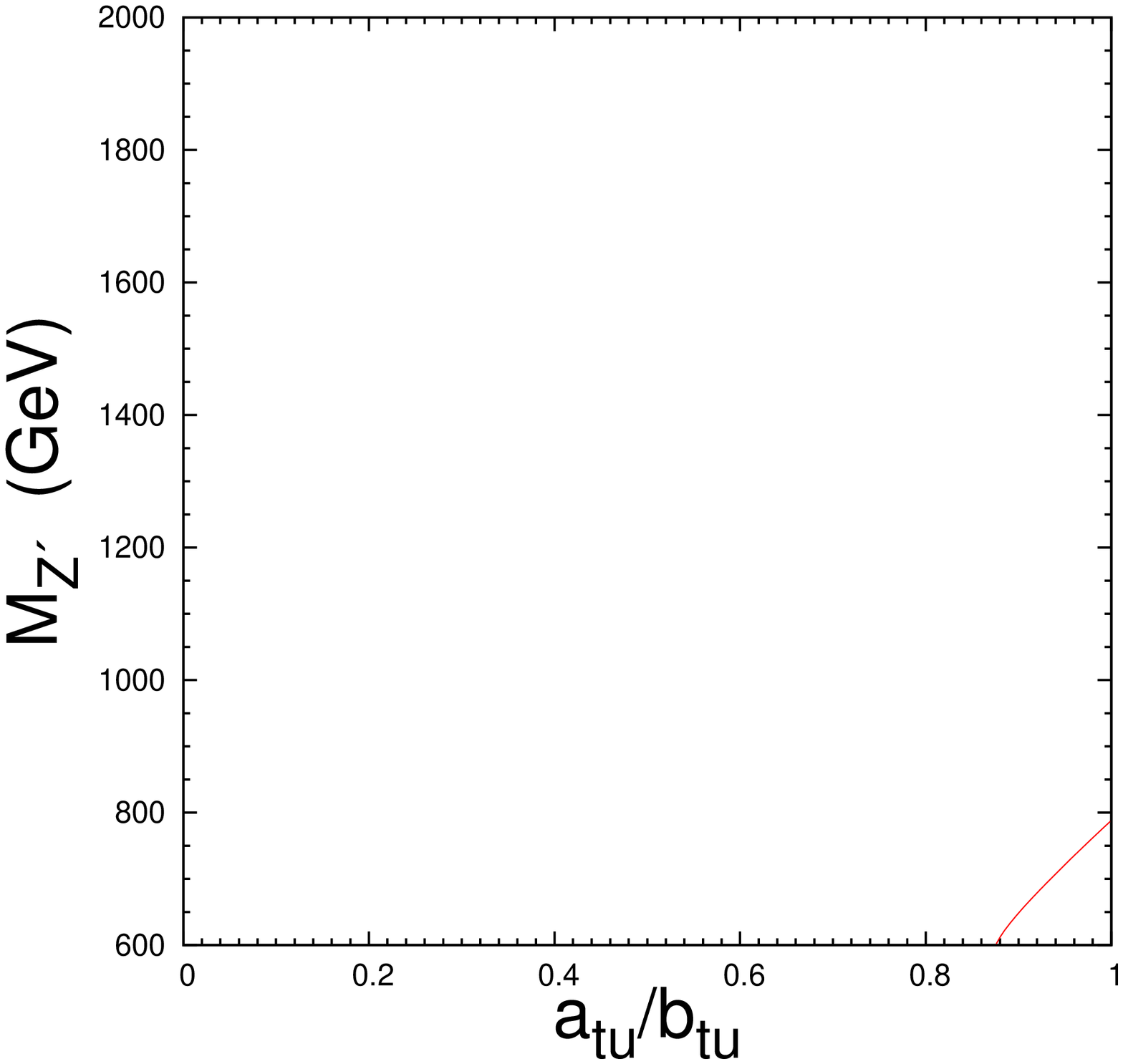}
\includegraphics[angle=-90,width=.49\textwidth]{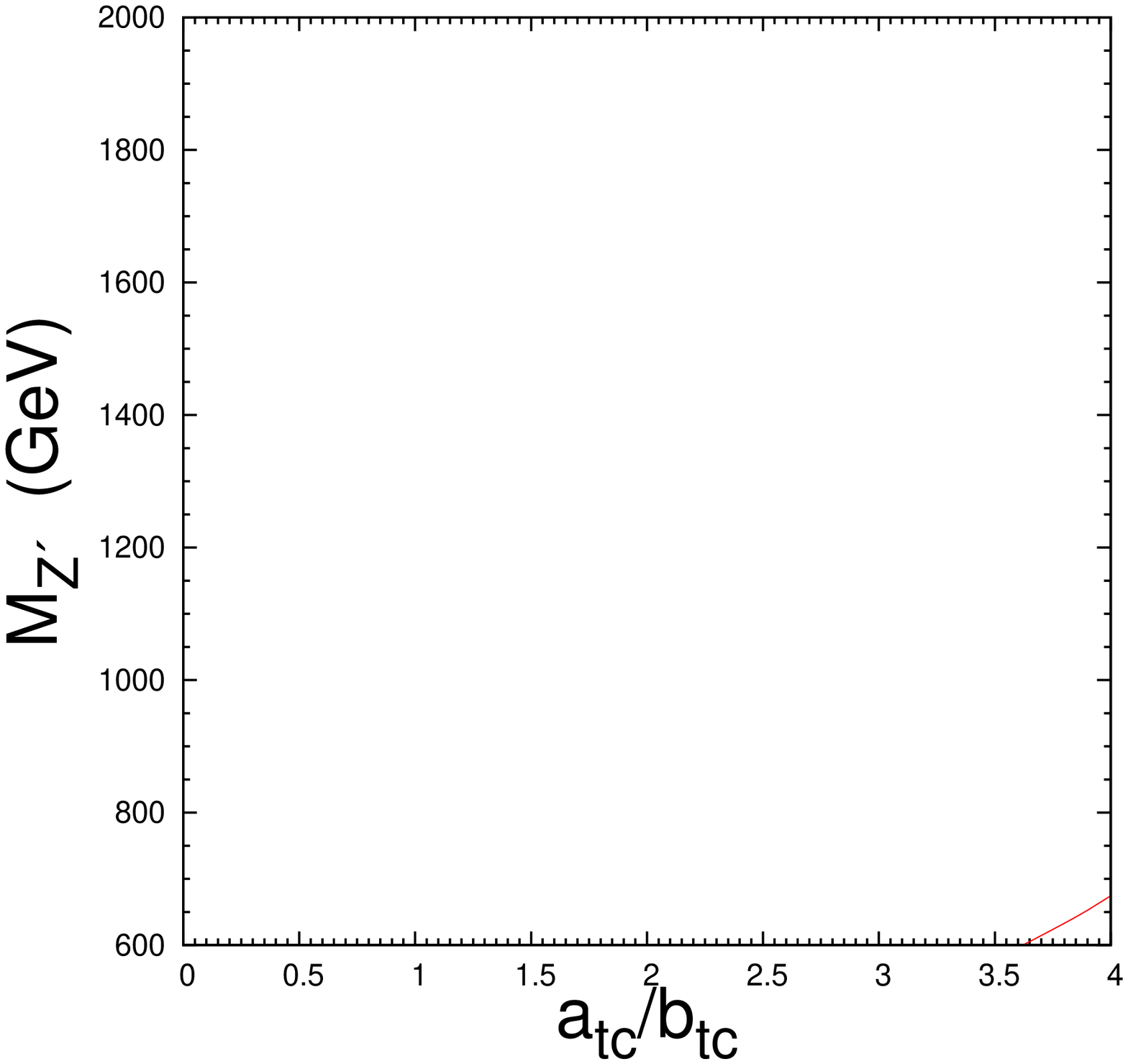}
\caption{The region to the right of the curve shows the parameter space accessible at the LHC in $Z'\to t\bar{t}$ detection mode for $\sqrt{s} = 14$ TeV and $\int {\cal L} dt = 100 fb^{-1}$ after the cuts discussed in the text.}
\label{f:gtt}
\end{figure}

We now present our results in  Tables [\ref{t:prelimc}-\ref{t:finalc}] and Figures [\ref{f:gmm}-\ref{f:gtt}] which give the minimum value of the particular coupling for a given $Z^\prime$ mass that results in 10 signal events per 100 fb$^{-1}$.  For cases (a) and (b) above the cuts reduce the background to 17 events per 100 fb$^{-1}$ so our tables/figures correspond to a signal significance $S/\sqrt{B} \sim 2.4$. For cases (c) and (d) the cuts completely remove the background at the level of our study.

%%%%%%%%%%%% 
\begin{table}[h]
\centering
\begin{tabular}{|c|c|c|c|c|}\hline
$Z'$ Decay Mode& $a_{tu}$ or $b_{tu}$  & $a_{tc}$ or $b_{tc}$  & $a_{tu}$ or $b_{tu}$ & $a_{tc}$ or $b_{tc}$  \\
& $M_{Z^\prime}=0.6$~TeV & $M_{Z^\prime}=0.6$~TeV & $M_{Z^\prime}=1 $~TeV &$M_{Z^\prime}=1$~TeV \\\hline
$Z'\to \mu^+\mu^-$&0.014&0.055&0.038&0.18\\
$Z'\to \tau^+\tau^-$&0.085&0.32&0.21&0.96\\
$Z'\to b \bar{b}$&0.014&0.056&0.038&0.21\\
$Z'\to t\bar{t}$&0.13&0.45&0.37&1.6\\
\hline
\end{tabular}
\caption{Lowest value of the corresponding couplings that yields 10 events per 100 fb$^{-1}$ before any cuts have been applied.}
\label{t:prelimc}
\end{table}

\begin{table}[h]
\centering
\begin{tabular}{|c|c|c|c|c|}\hline
$Z'$ Decay Mode& $a_{tu}$ or $b_{tu}$  & $a_{tc}$ or $b_{tc}$  & $a_{tu}$ or $b_{tu}$ & $a_{tc}$ or $b_{tc}$  \\
& $M_{Z^\prime}=0.6$~TeV & $M_{Z^\prime}=0.6$~TeV & $M_{Z^\prime}=1 $~TeV &$M_{Z^\prime}=1$~TeV \\\hline
$Z'\to \mu^+\mu^-$&0.025&0.097&0.057&0.27\\
$Z'\to \tau^+\tau^-$&0.18&0.78&0.36&1.7\\
$Z'\to b \bar{b}$&0.025&0.096&0.058&0.28\\
$Z'\to t\bar{t}$&0.92&3.6&1.2&5.7\\
\hline
\end{tabular}
\caption{Lowest value of the corresponding couplings that yields 10 events per 100 fb$^{-1}$ after all the cuts have been applied.}
\label{t:finalc}
\end{table}

Table~\ref{t:prelimc} shows the results before any cuts are applied whereas Table~\ref{t:finalc} presents the corresponding numbers after all the cuts have been applied. All bounds are obtained assuming a 100\% branching ratio ${\cal B}$ for the $Z^\prime$ to decay to the corresponding mode. The numbers in the Tables thus scale with $\sqrt{\cal {B}}$. The salient features of these Tables are:

\begin{enumerate}[a)]
\item The difference between the numbers in the $Z^\prime \to \mu^+\mu^-$ row in Table~\ref{t:prelimc} and the preliminary numbers in Table~\ref{tabrough} simply reflects the decrease in statistics from the requirement of identifying the top through its semileptonic decay (to $\mu$).

\item Similarly, the worsening in the limits in going to the $Z^\prime \to \tau^+\tau^-$ and/or the $Z^\prime \to t\bar{t}$ modes simply reflects the additional loss of statistics from the two $\tau ({\rm ~or~} t) \to \mu \cdots$ branching ratios. In this sense the numbers in the Table are very conservative as we have chosen only the cleanest decay mode for both the $\tau$-lepton and top-quark cases. For example, we could assume instead that taus can be tagged  with identification efficiency ($\sim 45 \%$) as reported in Ref.~\cite{Aad:2009wy}, the limits from $Z'\to \tau^+ \tau^-$ can be improved by a factor of 3 or so. 

\item The background can be essentially eliminated with the sets of cuts we suggest and the loss in signal sensitivity is relatively small (factors of 2-3), as can be seen by comparing Tables~\ref{t:prelimc}~and~\ref{t:finalc}. This result is encouraging although additional studies at the detector level are still needed.

\end{enumerate}

Tables~\ref{t:prelimc}~and~\ref{t:finalc} illustrate the sensitivity limits for two values of $M_{Z^\prime}$. The corresponding limits as a function of $M_{Z^\prime}$ are presented in Figures~\ref{f:gmm}, \ref{f:gll}, \ref{f:gbb},  and \ref{f:gtt} for the case of $Z^\prime t u$ and $Z^\prime t c$ couplings for four different decay modes of the $Z^\prime$.

%--------------------------------------------------------------------
%--------------------------------------------------------------------

%--------------------------------------------------------------------
%--------------------------------------------------------------------

%--------------------------------------------------------------------
%--------------------------------------------------------------------

%--------------------------------------------------------------------
%--------------------------------------------------------------------

There is one additional background that we have not mentioned because it is due to the new physics itself. Flavor diagonal couplings of a non-universal $Z^\prime$ can give rise to the same final state we considered in this paper. An example is single top production in the $W b\to t$  channel followed by a flavor diagonal $Z^\prime$ coupling to top as studied in Ref.~\cite{Han:2004zh}. This particular background is model dependent and could be large for models in which the $Z^\prime t\bar t$ couplings are very large \cite{modelone}. However, being due to the new particle itself, it is something to study after discovery of a $Z^\prime$ as also suggested in Ref.~\cite{Chen:2008za}.

\section{Conclusions}

We have studied the bounds that can be placed by the LHC on flavor changing couplings of a new $Z^\prime$ boson to top-charm quarks or top-up quarks using the single top production in association with the $Z^\prime$ process. We have first presented preliminary results derived from counting the number of $Z^\prime t+Z^\prime\bar{t}$ events. These results indicate that integrated luminosities of a few hundred fb$^{-1}$ at 14~TeV are needed to reach the interesting region of parameter space vis-a-vis low energy constraints on FCNC and in particular $D-\bar{D}$ mixing.

We have then extended our study to make it more realistic by considering the effects of SM background. To this end we looked for SM events with the same topologies as the signal events in four different decay channels of the $Z^\prime$ (to $\mu^+\mu-, \tau^+\tau^-, b\bar{b}, {\rm~and~} t\bar{t}$) and used appropriate cuts to reduce or eliminate the background. At this level of analysis we found that it is possible to manage these backgrounds at a modest cost to the signal. Our analysis for the $\tau$-lepton and top-quark modes is overly conservative as we assumed they could only be identified in the semileptonic (to muons) mode. Improvements on this would drive the region that can be probed towards the ideal region shown in Figure~\ref{f:prelim}.

\begin{acknowledgments}

This work was supported in part by DOE under contract number DE-FG02-01ER41155.

\end{acknowledgments}

\end{document}